\documentclass{article}

\title{Charged complexes at the surface of liquid helium}
\author{A.M. Dyugaev, P.D. Grigoriev, P. Wyder \\
L.D.Landau Institute for Theoretical Physics, Chernogolovka, Russia\\
Grenoble High Magnetic Field Laboratory, MPI-FKF and CNRS,
Grenoble, France}

\begin{document}
\maketitle

(In honor of Professor Jozef T. L. Devreese's 65th birthday)

\medskip

\begin{abstract}
Charged clusters in liquid helium in an external electric field
form a two-dimensional system below the helium surface. This 2D
system undergoes a phase transition from a liquid to a Wigner
crystal at rather high temperatures. Contrary to the electron
Wigner crystal, the Wigner lattice of charged clusters can be
detected directly.
\end{abstract}

\textbf{1}. The properties of charged particles on the surface and
in the bulk of liquid helium have been extensively studied for
more than 30 years \cite{Shikin}. An electron on the helium
surface has very high mobility due to the high purity of helium at
low temperatures, when all outside particles quit the quantum
liquid. In the bulk of helium an electron has a low mobility since
it forms a bubble of a large radius $\sim 17 $\AA . Negative ions
as $H^{-}$ or $O_{2}^{-} $ in the bulk of liquid helium also
attracted some attention \cite{VKS}. In \cite{NI} the surface
states of negative ions of large radius (as $Ca^{-}$ and $Ba^{-}$)
were predicted and studied. One interesting point is that the
negative ions $Hg^{-}, \ Be^{-}, \ Zn^{-}, \ Cd^{-} $ which do not
exist in vacuum may exist in liquid helium. The atoms of these
elements have a very great polarizability, and an electron may
almost form a bound state with these atoms. In helium an electron
is already confined so that these bound states realize.

There are several experimental methods to introduce an outer
particle into liquid helium. Atoms, molecules and small clusters
are created inside helium by the laser ablation. Larger clusters
of radius $R>50 $\AA \ are instilled into helium through its
surface \cite{GorMez}. At high concentration of these clusters in
helium an iceberg is formed. This was studied experimentally in
\cite{Mezhov} by injecting clusters of $\sim 10^{3} $ molecules
into liquid helium. Large clusters of air or hydrogen were
observed as a fog, slowly falling on the bottom of the vessel at
speed of $ 10^{-2} cm/sec $. Below the $\lambda $-point of helium
the clusters coagulate in the flakes of a ''snowfall''
\cite{Mezhov}.

In the present paper we study the properties of large charged
clusters in liquid helium. We show that in strong electric field
one can create a two-dimensional charged system of heavy particles
below the surface of the liquid. This system may give an
opportunity to detect directly its Wigner crystallization as the
temperatures decreases.

\medskip

\textbf{2}. Keeping in mind recent experiments with the crystals
of $H_{2}O$ in helium \cite{Mezhov}, we shall simulate the
impurity particle as a macroscopic dielectric sphere of radius
$R$. As is well known \cite{LLESS}, the interaction $ V_{eB}(r)$
of an electron with such a sphere at large distance $r>R$ has the
form
\begin{equation}
V_{eB}=\frac{e\vec{E}_{ex}\vec{r} }
{r^{3}}R_{*}^{3}-\frac{e^{2}R_{*}^{3}}{2%
r ^{4}},  \label{Hint}
\end{equation}
where
\begin{equation}
  R_*^{3}\equiv R^{3}\frac{\varepsilon _{B}-\varepsilon
_{h}}{\varepsilon _{B}+2\varepsilon _{h}}. \label{R*}
\end{equation}
The first term in (\ref{Hint}) is the interaction of an electron
with the dipole moment of the cluster $\vec{d}=\vec{E}_{ex}R_{*}
^{3}$ created by the external field $\vec{E}_{ex}$. The second
term in (\ref{Hint}) is due to the polarization of the cluster by
the electron. As can be seen from (\ref{R*}) the effective size of
the cluster $R_*$ is equal to $R$ if its dielectric constant
$\varepsilon _{B}\gg 1$. Note that the dielectric constant of
helium is $\varepsilon _{h}=1.054$. If in helium there are some
electrons and impurities, the interaction (\ref{Hint}) leads to
the formation of charged clusters with large binding energy
$E_{Be}\sim 0.1 - 1 eV$. The electric field $E_{ex}$ presses these
heavy charges to the surface of helium, where the clusters
organize into a two-dimensional system. The optimal distance
$z^{opt}_{B}$ of these heavy charged particles to the surface is
determined by the minimum of the potential energy:
\begin{equation}
V_{cl}(z_{B})=eE_{ex}z_{B}+ \frac{e^{2}}{z_{B}} 2\nu_1 - Mgz_{B} +
\frac{(\vec{E}_{ex}R_{*} ^{3})^{2}}{z_{B}^{3}}\nu_1 + V_{vdW},
\label{VzB}
\end{equation}
where
$$
\nu_1 \equiv \frac{1}{8}\frac{\varepsilon _{h}-1}{\varepsilon
_{h}+1} \approx 1/300.
$$
The first term in (\ref{VzB}) is due to the external field. The
second term is the image potential of a charge near the surface.
The third gravitational term is important only for very big
clusters of radius $R >10^{4}$\AA ; usually, one can neglect the
gravitational term. The fourth term in (\ref{VzB}) is the image
potential of the electric dipole moment of the cluster.
Rigorously, the dipole moment of the cluster $\vec{d}$ is
generated both by the external field $\vec{E}_{ex}$ and by the
field from the attracted electron. However, the image potential
due to electron-generated dipole moment of the cluster is always
smaller than the second term in (\ref{VzB}) by a factor of
$[(\varepsilon _{B}-1)R/(\varepsilon _{B}+1)z_B ]^{2} /2 < 1/2$.
Therefore, for the estimation of $z_B$ we shall take into account
only the dipole moment $\vec{d}=\vec{E}_{ex}R_{*} ^{3}$ generated
by the external field. The last term in (\ref{VzB}) is the van der
Waals potential. This term prevents the cluster to emerge from
helium. It is $\propto 1/z_B^{3} $ and becomes important only when
the cluster is very close to the surface.

Now we have to separate two different cases. First we consider
small clusters when $e/R^{2} \gg E_{ex}$ or $R\ll
\sqrt{e/E_{ex}}\approx 700$\AA \ for a field $E_{ex}=3000V/cm$.
The gravitational field, the dipole image potential and the van
der Waals potential are then much weaker than the first two terms
in (\ref{VzB}). Therefore, to find the optimal $z_B$ we shall
consider only these two (pure electronic) terms, and $z_B$ now
means the z-coordinate of the bound electron. Differentiating
$V_{cl}(z_{B})$ with respect to $z_{B}$ we find
\begin{equation}
z_{B} ^{opt}=\sqrt{ 2\nu_1 e/E_{ex}}. \label{zopt}
\end{equation}
For a field of $E_{ex}=3000V/cm$ we get $z_{B} ^{opt}\approx 60
$\AA . If $z_{B} ^{opt}<R$ the cluster will be close to the
surface but it will not emerge from helium because of the
short-range van der Waals attraction forces between cluster and
helium atoms.

The opposite case of a large cluster ($e/ R^{2} \ll E_{ex}$ or
$R^{2}\gg (700 $\AA $)^{2}$) is even more interesting. Now we have
to consider also the dipole-dipole image potential of the cluster
as it increases as $R^{6}$. The optimal distance $z_{B} ^{opt}$ is
now determined by the quadratic equation
\begin{equation}
\frac{d V_{cl}}{d z_{B}}=eE_{ex}- \frac{e^{2}}{z_{B}^{2}} 2\nu_1 -
  3\frac{(\vec{E}_{ex}R_{*} ^{3})^{2}}{z_{B}^4}\nu_1 = 0, \label{VzEq}
\end{equation}
which gives
\begin{equation}
z_{B} ^{opt}=\sqrt{ \frac{\nu_1 e}{E_{ex}}} \left[ 1+\sqrt{
1+\frac{3}{\nu_1 } \left( \frac{E_{ex}R_{*} ^{2}}{e}\right) ^{3}}
\right] ^{1/2} \label{z0big}
\end{equation}
At $E_{ex}R_{*} ^{2}/{e}\gg 1$ this becomes
\begin{equation}
z_{B} ^{opt} \approx \sqrt{ \frac{\nu_1 e}{E_{ex}}}
\left(\frac{3}{\nu_1 }\right) ^{1/4} \left(\frac{E_{ex}R_{*}
^{2}}{e}\right) ^{3/4}.  \label{z0big1}
\end{equation}
Now the optimal distance $z_{B} ^{opt} \propto (E_{ex})^{1/4}
R^{3/2}$, and one can reach $z_{B} ^{opt}>R$ by increasing
$E_{ex}$ and $R$. At $\varepsilon _{B}=80 \approx \infty $ and
$E_{ex}=5000 V/cm$ this condition is fulfilled only for rather
huge clusters $R\geq 5000$\AA $=0.5\mu m $.

Above we have argued the possibility of realization of a 2D system
of charged clusters below a helium surface. The dipole moment and
electric charge of each cluster allow to fix the cluster at a
macroscopic distance $z_0>R$ below the helium surface. At low
helium temperatures any thermal fluctuations of the positions of
these heavy charges are small. Let us now consider the Wigner
crystallization of these charges.

The interaction of two charged clusters separated by distance $a$
is described by the potential energy
\begin{equation}
V_{int}(a)= \frac{e^{2}}{a}+\frac{(\vec{E}_{ex}R_{*}
^{3})^{2}}{a^{3}}\equiv V_{e}+V_{d}.
  \label{Va}
\end{equation}
The dipole-dipole part $V_{d}$ of the interaction between clusters
becomes dominant ($V_{d}>V_{e}$) if $a e < E_{ex}R_{*} ^{3}$. At
$E_{ex}=5\cdot 10^{3} V/cm$, $R_{*}=5000$\AA \ and $a =10R_{*}$
(that corresponds to the average cluster density
$n_s=1/a^{2}=4\cdot 10^{6} cm^{-2}$) one gets
\begin{equation}
V_{e}=3K; \ \ \ V_{d}=300K.
  \label{VEst}
\end{equation}
Accepting the empirical criterion $V(a)>100k_{B}T$ of melting of
2D Wigner crystal we come to the conclusion of the possibility to
observe the  liquid-crystal phase transition at rather high
temperature $T\sim 3K$. The oddity of the situation in hand is
that the crystal lattice is formed by the short-range dipole
forces ($V_d \gg V_e$), and the long-range Coulomb interaction
appears in the existence of long-wave plasma oscillations. An
observation of these effects should be possible by using the
method first applied 50 years ago \cite{JUSSR}. The neutral
clusters come to helium through its surface and go down slowly to
the bottom of the helium vessel where the electron source is
maintained. The electron from the injector are drawn up by the
electric field. In the bulk of helium some of the clusters form a
bound with one electron and form a 2D charged system below the
surface. Other clusters glue to the bottom of the vessel by the
van der Waals forces. The excess electrons will finally make a
quantum tunnel transition through helium surface to the vacuum
where the electrons have a lower energy. The energy of a big
charged clusters is, on the contrary, lower inside the helium
because of the short-range van der Waals attraction of the cluster
to the atoms of liquid. This gain of the  van der Waals energy is
roughly proportional to the surface of cluster.

To study the stability of the 2D charged system of clusters one
has to consider also their van der Waals attraction to each other
at the mean distance $a$. At large distance, the van der Waals
potential energy of two atoms is \cite{LL9}
\begin{equation}
V_{vdW}(a)= \frac{23\hbar c }{4 \pi a^{7} } \beta^{2}
  \label{VavdW}
\end{equation}
where $\beta$ is the atomic polarizability of clusters. It is
related to the dielectric constant $\varepsilon_B=1+4\pi\beta n$,
where $n$ is the density of the medium of the clusters.
Integration of (\ref{VavdW}) over the cluster volume gives
\begin{equation}
V_{BB}(a)= \frac{23\hbar c R^{6}}{36 \pi a^{7} }
(\varepsilon_B-\varepsilon_H)^{2}
  \label{VBBvdW}
\end{equation}
For a cluster consisting of the molecules of $H_2O$ with
$\varepsilon_B \approx 80$, for the parameters $R=5000$\AA \ and
$a=10R$ (the same as in (\ref{VEst})) from (\ref{VavdW}), we get
an estimate of $V_{BB}(a)=0.6K$. Since $V_{BB}(a)\ll V_{d}(a)$,
there is a wide range of parameters where the 2D system under
study is stable against coagulation (sticking together) of
clusters.

It is much easier to detect the Wigner lattice of huge charged
clusters than that of electrons. The Wigner lattice of electrons
allows only an indirect method for experimental studies \cite{GA}.
On the other hand, the Wigner crystallization of charged clusters
can be detected directly by neutron or X-ray scattering. If the
charged clusters are very huge ($R\gg 5000$\AA) they can be
visually observed.

In closing we note that the considered effects are not very
sensitive to the type of quantum liquid. They could also be
observed in liquid hydrogen, where $\varepsilon_H=1.28$ and the
parameter $\nu_1=1.5\cdot 10^{-2}$.

\medskip

The work was supported by INTAS-01-0791 and RFBR No 00-02-17729.

\appendix

\bigskip

{\large Appendix.}

\medskip

In this appendix we study the structure of the charged cluster in
more detail. The attraction of electron to the cluster (at large
distance given by eq. (\ref{Hint})) is quite strong so that the
binding energy of the electron is large compared to the potential
energy (\ref{VzB}). The electron is localized at one side of the
cluster (if the electron was uniformly distributed along the
surface of cluster it would not gain any polarization energy). The
mean distance of the electron from the cluster is always much
smaller than the cluster radius: $z_e\ll R$. To estimate the
binding energy of the electron and the dipole moment of the
charged cluster we shall consider the surface of the cluster near
the electron to be flat (the problem of a point-like charge near
dielectric sphere cannot be solved in finite form \cite{LLESS}).
Then the polarization of the cluster by the electron can be taken
into account via the image potential
$$
V_{eB}=\frac{e^{2}}{8 z_e}\frac{\varepsilon _{B}-1}{\varepsilon
_{B}+1}
$$
which determines the average distance $\overline{z_e}$. This is a
one-dimensional Coulomb potential similar to that of an electron
above helium surface without external field. Above the helium
surface the electron levitates on a distance of $\overline{z^h _e}
= 114 $\AA \cite{Shikin}. In the case of an electron near the
dielectric cluster we have
$$\overline{z _e} =
\frac{\varepsilon _{h}-1}{\varepsilon _{h}+1} \left(
\frac{\varepsilon _{B}-1}{\varepsilon _{B}+1} \right) ^{-1}
\overline{z^h _e}.
$$ For $\varepsilon _{B}=80 \approx \infty$, $
\overline{z _e} = 0.027 \overline{z^h _e} \approx 3$\AA $ \ll
R_{B}$. The corresponding electron-generated dipole moment of the
cluster is $d_{e-g} \approx R_{B} e(\varepsilon
_{B}-1)/(\varepsilon _{B}+1) = eR^{3} _{*}/R^{2}$. It becomes
equal to the dipole moment $\vec{d}_{ex}=\vec{E}_{ex}R_{*}^{3}$
generated by the external field at $E_{ex}=e/R^{2}\approx 1500
V/cm$ for $R=1000$\AA .

The coordinate $z^{opt}_{B}$ determined from the potential
(\ref{VzB}) is actually the z-coordinate of the center of charge
(not the center of the cluster). $z^{opt}_{B}$ coincides with the
center of cluster only for $\varepsilon _{B}\rightarrow \infty $.
If $z^{opt}_{B}$ obtained from (\ref{VzB}) is equal or less than
the cluster radius $R$, one has to take into account the
short-range van der Waals potential, which prevents the cluster
from emerging from helium.

\bigskip

Figure caption:  The schematic view of the charged clusters below
helium surface.

\end{document}